\documentclass[a4paper,aps,prd,10pt,preprintnumbers,showpacs,superscriptaddress,nofootinbib,amsmath,amssymb,floatfix
,twocolumn
]{revtex4-1}

\usepackage{enumitem}
\usepackage{graphicx}
\usepackage[utf8]{inputenc}
\usepackage[T1]{fontenc}
\usepackage{cmap}
\usepackage{graphicx}
\usepackage{dcolumn}
\usepackage{bm}
\usepackage{bbm}
\usepackage{amsfonts}
\usepackage{amsmath}
\usepackage{enumitem}
\usepackage{lipsum}
\usepackage[bottom]{footmisc}
\usepackage{cmap}
\usepackage{subcaption}
\usepackage{braket}
\usepackage{braket}
\usepackage[table]{xcolor}
\usepackage{amsmath}
\usepackage{hyperref}
\usepackage{cleveref}
\usepackage{fmtcount}
\usepackage{dblfnote}
\usepackage{float}
\usepackage{amssymb}
\usepackage{natbib}
\usepackage[english]{babel}
\usepackage{blindtext}

\begin{document}

\title{Shadows, greybody factors, emission rate, topological charge, and phase transitions for a charged black hole with a Kalb--Ramond field background}
\author{F. Hosseinifar}
\email{f.hoseinifar94@gmail.com}
\affiliation{Department   of   Physics,   University   of   Hradec   Kr\'{a}lov\'{e}, Rokitansk\'{e}ho   62, 500   03   Hradec   Kr\'{a}lov\'{e},   Czechia}
% % %
\author{A. A. Araújo Filho}
\email{dilto@fisica.ufc.br}
\affiliation{Departamento de Física, Universidade Federal da Paraíba, Caixa Postal 5008, 58051-970, João Pessoa, Paraíba, Brazil}
% % %
\author{M. Y. Zhang}
\email{gs.myzhang21@gzu.edu.cn}
\affiliation{School of Mathematics and Statistics, Guizhou University, Guiyang, 550025, China}
% % %
\author{H. Chen}
\email{haochen1249@yeah.net}
\affiliation{School of Physics and Electronic Science, Zunyi Normal University, Zunyi 563006,PR China}
% % %
\author{H. Hassanabadi}
\email{hha1349@gmail.com}
\affiliation{Department   of   Physics,   University   of   Hradec   Kr\'{a}lov\'{e}, Rokitansk\'{e}ho   62, 500   03   Hradec   Kr\'{a}lov\'{e},   Czechia}
%%%%%%%%%%%%%%%%%%%%%%%%%%%%%%%%%%%%%%%%%%%%%%%%%%%%%%%%%%%%%%%%%%%%%%%%%%%%%%%%%%%%%%%%%%%%%%%%%%%%%%%%%%%%%%%%%%%%%%%%%%%%%%%%%%%%%%%%%%%%%%%%%%%%%%%%%%%%%%%%%%%%%%%%%%%%%%%%%%%%%%%%%%%%%%%%%%%%%%%%%%%%%%%%%%%%%%%%%%%%%%%%%%%%%%%%%%%%%%%%%%%%%%%%%%%%%%%%%%%%%%%%%%%%%%%%%%%%%%%%%%%%%%%%%%%%%%%%%%%%%%%%%%%%%%%%%%%%%%%%%%%%%%%%%%%%%%%%

\begin{abstract}

In this work, we investigate a spherically symmetric charged black hole in the presence of a Kalb--Ramond field background. We calculate the photon sphere and shadow radii and, corroborating our results, we constrain them from observational data from the Event Horizon Telescope (EHT), particularly focusing on the shadow images of Sagittarius $A^{*}$. Additionally, we analyze the \textit{greybody} factors, emission rate, and partial absorption cross section. We also examine the topological charge and its application to the deflection angle. Finally, we conduct the analysis of the heat capacity and phase transitions.

\end{abstract}
\maketitle

%%%%%%%%%%%%%%%%%%%%%%%%%%%%%%%%%%%%%%%%%%%%%%%%%%%%%%%%%%%%%%%%%%%%%%%%%%%%%%%%%%%%%%%%%%%%%%%%%%%%%%%%%%%%%%%%%%%%%%%%%%%%%%%%%%%%%%%%%%%%%%%%%%%%%%%%%%%%%%%%%%%%%%%%%%%%%%%%%%%%%%%%%%%%%%%%%%%%%%%%%%%%%%%%%%%%%%%%%%%%%%%%%%%%%%%%%%%%%%%%%%%%%%%%%%%%%%%%%%%%%%%%%%%%%%%%%%%%%%%%%%%%%%%%%%%%%%%%%%%%%%%%%%%%%%%%%%%%%%%%%%%%%%%%%%%%%%%%%%%%%%%%%%%%%%%%%%%%%%%%%%%%%%%%%%%%%%%%%%%%%%%%%%%%%%%%%%%%%%%%%%%%%%%%%%%%%%%%%%%%%%%%%%%%%%%%%%%%%%%%%%%%%%%%%%%%%%%%%%%%%%%%%%%%%%%%%%%%%%%%%%%%%%

\section{Introduction}\label{Sec1}

Lorentz symmetry breaking \cite{bluhm2005spontaneous,bluhm2021gravity,kostelecky2011matter,kkos,tasson2014} is a remarkable concept in modern physics and has become a significant area of research across various theoretical frameworks. One approach to studying Lorentz symmetry breaking involves introducing the antisymmetric tensor field known as the Kalb--Ramond (KR) field \cite{altschul2010lorentz}, which originates from string theory \cite{kalb1974classical}.

The aspects of black hole (BH) physics within the framework of KR gravity have been explored extensively \cite{3,2,1}. When the KR field is coupled with gravity, it can induce spontaneous Lorentz symmetry breaking \cite{altschul2010lorentz}. This phenomenon was examined by deriving an exact solution for a static, spherically symmetric BH configuration \cite{5}. Building on this, subsequent studies have investigated the behavior of massive and massless particles near KR black holes \cite{6}, as well as the gravitational deflection of light and the shadows cast by rotating black holes \cite{7}. Furthermore, considerable attention has been given to the discovery of gravitational waves and their spectrum within the realm of Lorentz symmetry breaking \cite{2,8,9,10,11}.

On the other hand, the effects of Lorentz symmetry violation on electrically charged black holes have been studied in KR gravity as well.
For a static, spherically symmetric spacetime, the metric can be written as:
\[
\mathrm{d}s^2=-f(r)\mathrm{d}t^2+\frac{1}{f(r)}\mathrm{d}r^2+h(r) \mathrm{d}\Omega^2\label{ds2},
\]
where, as introduced in Refs. \cite{3}, we have:
\begin{equation}
\label{fr}    
f(r)=\frac{1}{1-\bar{l}}-\frac{2M}{r}+\frac{Q}{(1-\bar{l})^2 r^2}.
\end{equation}
Here, \(M\) is the BH mass, \(Q\) is the electric charge, and \(\bar{l}\) is the Lorentz--violating parameter and \(h(r)=r^2\). In this context, the gravitational lensing was analyzed via weak and strong deflection limits as well as the time delay\cite{17}. Ref. \cite{18} investigates a Lorentz--violating metric that includes an effective cosmological constant, while Ref. \cite{19} introduces another form of Lorentz violation and studies the metric's dynamical properties. On the other hand, a variety of aspects by considering such a metric with zero electric charge though was discussed in Refs. \cite{13,14,15,16,araujo2024exploring}.

This study explores a spherically symmetric charged black hole within a Kalb--Ramond field background. The photon sphere and shadow radii are determined and validated against observational data from the Event Horizon Telescope (EHT), with a particular focus on the shadow images of Sagittarius $A^{*}$. Moreover, the \textit{greybody} factors, emission rate, and partial absorption cross section are examined. The topological charge and its influence on the deflection angle are also investigated. Finally, the heat capacity and phase transitions are analyzed.

This work is organized as follows: in Sec. \ref{Sec2}, we perform the shadow radii and their respective constrains based on observational data from Sagittarius $A^{*}$; in Sec. \ref{Sec3}, we study the heat capacity for the system. In Sec. \ref{Sec4}, we discuss about the \textit{greybody} bounds, its relevant emission power, and the absorption cross section. Secs. \ref{Sec5}, \ref{Sec6} and \ref{Sec7}, are related to the topological aspects of the BH and its charge, and finally in Sec. \ref{Sec9}, we give the final concluding remarks. 

%%%%%%%%%%%%%%%%%%%%%%%%%%%%%%%%%%%%%%%%%%%%%%%%%%%%%%%%%%%%%%%%%%%%%%%%%%%%%%%%%%%%%%%%%%%%%%%%%%%%%%%%%%%%%%%%%%%%%%%%%%%%%%%%%%%%%%%%%%%%%%%%%%%%%%%%%%%%%%%%%%%%%%%%%%%%%%%%%%%%%%%%%%%%%%%%%%%%%%%%%%%%%%%%%%%%%%%%%%%%%%%%%%%%%%%%%%%%%%%%%%%%%%%%%%%%%%%%%%%%%%%%%%%%%%%%%%%%%%%%%%%%%%%%%%%%%%%%%%%%%%%%%%%%%%%%%%%%%%%%%%%%%%%%%%%%%%%%%%%%%%%%%%%%%%%%%%%%%%%%%%%%%%%%%%%%%%%%%%%%%%%%%%%%%%%%%%%%%%%%%%%%%%%%%%%%%%%%%%%%%%%%%%%%%%%%%%%%%%%%%%%%%%%%%%%%%%%%%%%%%%%%%%%%%%%%%%%%%%%%%%%%%%%%%%%%%%%%%%%%%%%%%%%%%%%%%%%%%%%%

\section{Shadows}\label{Sec2}

\begin{figure}[ht!]
\centering
	\begin{subfigure}[b]{0.5\textwidth}
	\centering
	\includegraphics[scale = 0.48]{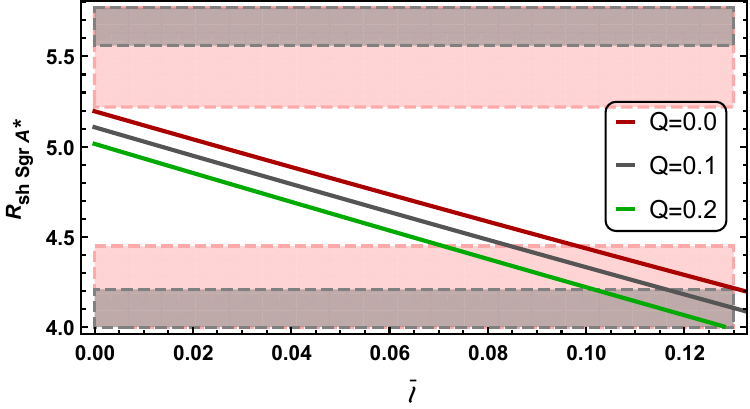} \hspace{-0.2cm}\\
%	\caption{}
    \end{subfigure}%
\hfill
\caption{Shadow radius versus parameter $\bar{l}$. Pink area is denied for $1\sigma$ and gray region is not acceptable for $2\sigma$.}
\label{fig:shadowl}
\end{figure}

\begin{figure}[ht!]
\centering
	\begin{subfigure}[b]{0.5\textwidth}
	\centering
	\includegraphics[scale = 0.48]{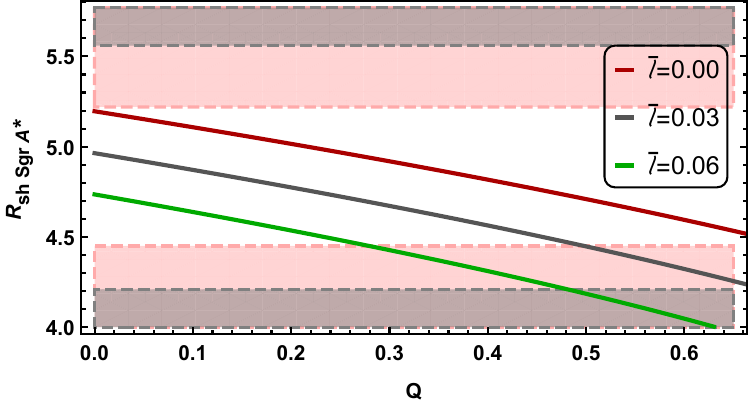} \hspace{-0.2cm}\\
%	\caption{}
    \end{subfigure}%
\hfill
\caption{Shadow radius versus parameter $Q$. Pink area is denied for $1\sigma$ and gray region is not acceptable for $2\sigma$.}
\label{fig:shadowQ}
\end{figure}

The motion of light can be found by using the Euler--Lagrange equation as follows
\begin{eqnarray}
\frac{\mathrm{d}}{\mathrm{d}\tau}\left(\frac{\partial \mathcal{L}}{\partial \dot{x}^{\mu}}\right)-\frac{\partial \mathcal{L}}{\partial x^{\mu}}=0,\,\,\,\,\, \mathcal{L}=\frac{1}{2}g_{\mu\nu}\dot{x}^{\mu}\dot{x}^{\nu}.
\end{eqnarray}
At $\theta=\pi/2$, defining two constant of motion as $L = h(r)\dot{\phi}$ and $E = f(r) \dot{t}$, and after algebraic manipulations, the equation of the effective potential \cite{chen2024quasi}
\begin{eqnarray}
V_{eff} = \frac{f(r)}{h(r)}\bigg(\frac{L^2}{E^2}-1\bigg)
\end{eqnarray}
is obtained. As the circular orbit imposed that $V_{eff}(r)=V^{'}_{eff}(r)=0$, though the photon behavior and specially the photon radius of a BH is calculated from \cite{29,30}
\begin{eqnarray}
r_{ph} \partial_r f(r)\bigg|_{r=r_{ph}}-2f(r_{ph})=0,
\end{eqnarray}
thus, photon radius is given by
\begin{eqnarray}
r_{ph}=-\frac{\sqrt{9 (\bar{l}-1)^4 M^2+8 (\bar{l}-1) Q}+3 (\bar{l}-1)^2 M}{2 (\bar{l}-1)}.\label{rph}
\end{eqnarray}
Due to the geometry, for an observer located at $r_o$ distance of the BH, shadow radius reads \cite{araujo2024dark}
\begin{eqnarray}
r_{sh}= \frac{r_{ph}}{\sqrt{f(r_{r_{ph}})}}\sqrt{f(r_{o})}.
\end{eqnarray}
The asymptotic form of Eq. (\ref{fr}), causes that in the limit of infinite radius, $f(r_o)$ becomes radius independent and shadow reduces to $r_{sh}= r_{ph}/f(r_{ph})$. The illustration of shadow radius curve by changing parameters is shown in  Figs. \ref{fig:shadowl} and \ref{fig:shadowQ}.
As can be seen, the downward behavior of the shadow radius according to the parameter  $\bar{l}$ ($Q$) 
is transferred to a lower radius with the increase of the $Q$ ($\bar{l}$) 
parameter. Due to observe shadow radius, Tabs. \ref{Table:rshl} and 
\ref{Table:rshQ}, indicates the permissible range for parameters $\bar{l}$ and $Q$, respectively.
\begin{table}[ht!]
		\caption{The allowed range of parameter $\bar{l}$, $\text{(upper bound, lower bound)}$, based on the obtained documentation to observations of Sgr $A^{*}$ BH, for different charges. Mass is set as $M=1$.}
		\centering
		\label{Table:rshl}
		\begin{tabular}{|c|c|c|}
		\hline
		 & \multicolumn{2}{c|}{Sgr $A^{*}$} \\
		 \cline{2-3}
		 $\bar{l}$ & $1\sigma$ & $2\sigma$ \\
		 \cline{2-3}
		 \hline
		 $Q=0.0$ & $\;(-,\,0.098)\;$ & $\;(-,\,0.131)\;$\\
		 \hline
		 $Q=0.1$ & $\;(-,\,0.084)\;$ & $\;(-,\,0.116)\;$\\
		\hline
%		\hspace{0.2cm}
		 $Q=0.2$ & $\;(-,\,0.071)\;$ & $\;(-,\,0.101)\;$\\
		 \hline
		\end{tabular}
	\end{table}
\begin{table}[ht!]
		\caption{The valid range of parameter $Q$, $\text{(upper bound, lower bound)}$, based on the obtained documentation to observations of Sgr $A^{*}$ BH, for different values of parameter $\bar{l}$, where mass is set as $M=1$.}
		\centering
		\label{Table:rshQ}
		\begin{tabular}{|c|c|c|}
		\hline
		 & \multicolumn{2}{c|}{Sgr $A^{*}$} \\
		 \cline{2-3}
		 $Q$ & $1\sigma$ & $2\sigma$ \\
		 \cline{2-3}
		 \hline
		 $\bar{l}=0.00$ & $\;(-,\,0.715)\;$ & $\;(-,\,0.881)\;$\\
		 \hline
		 $\bar{l}=0.03$ & $\;(-,\,0.498)\;$ & $\;(-,\,0.682)\;$\\
		\hline
		 $\bar{l}=0.06$ & $\;(-,\,0.279)\;$ & $\;(-,\,0.480)\;$\\
		 \hline
		\end{tabular}
	\end{table}

%%%%%%%%%%%%%%%%%%%%%%%%%%%%%%%%%%%%%%%%%%%%%%%%%%%%%%%%%%%%%%%%%%%%%%%%%%%%%%%%%%%%%%%%%%%%%%%%%%%%%%%%%%%%%%%%%%%%%%%%%%%%%%%%%%%%%%%%%%%%%%%%%%%%%%%%%%%%%%%%%%%%%%%%%%%%%%%%%%%%%%%%%%%%%%%%%%%%%%%%%%%%%%%%%%%%%%%%%%%%%%%%%%%%%%%%%%%%%%%%%%%%%%%%%%%%%%%%%%%%%%%%%%%%%%%%%%%%%%%%%%%%%%%%%%%%%%%%%%%%%%%%%%%%%%%%%%%%%%%%%%%%%%%%%%%%%%%%%%%%%%%%%%%%%%%%%%%%%%%%%%%%%%%%%%%%%%%%%%%%%%%%%%%%%%%%%%%%%%%%%%%%%%%%%%%%%%%%%%%%%%%%%%%%%%%%%%%%%%%%%%%%%%%%%%%%%%%%%%%%%%%%%%%%%%%%%%%%%%%%%%%%%%%%%%%%%%%%%%%%%%%%%%%%%%%%%%%%%%%%

\section{Heat Capacity}\label{Sec3}

\begin{figure}[ht!]
\centering
	\begin{subfigure}[b]{0.5\textwidth}
	\centering
	\includegraphics[scale = 0.48]{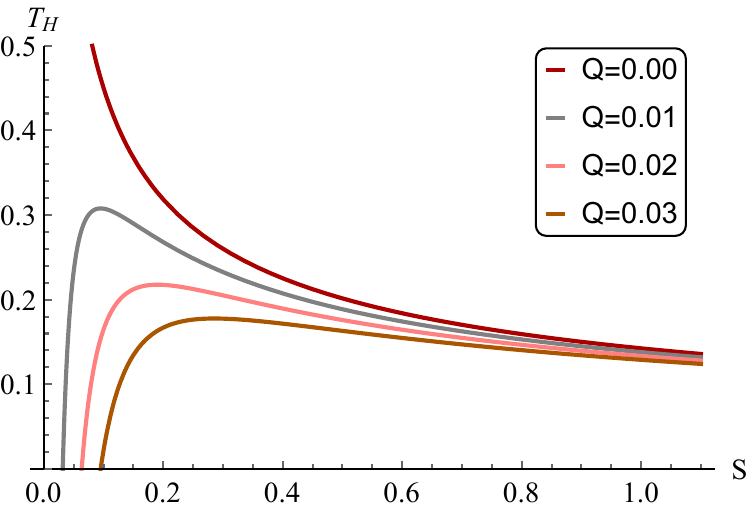} \hspace{-0.2cm}\\
%	\caption{}
    \end{subfigure}%
    \caption{The \textit{Hawking} temperature versus entropy. Parameter $\bar{l}$ is set as $\bar{l} = 0.01$.}
    \label{fig:TS}
\end{figure}

\begin{figure}[ht!]
\centering
	\begin{subfigure}[b]{0.5\textwidth}
	\centering
	\includegraphics[scale = 0.48]{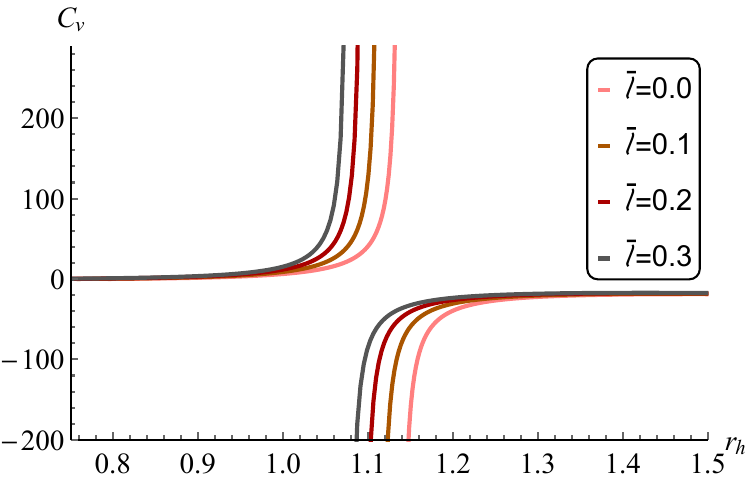} \hspace{-0.2cm}\\
%	\caption{}
    \end{subfigure}%
%\hfill
%	\begin{subfigure}[b]{0.5\textwidth}
%	\centering
%	\includegraphics[scale = 0.48]{cv3.pdf} \hspace{-0.2cm}\\
%%	\caption{}
%    \end{subfigure}%
\caption{The heat capacity versus $r_h$ by varying parameter $Q$. Mass is set as $M=1$.}
\label{fig:Cv}
\end{figure}

The thermodynamic properties and phase transitions of a BH are fundamental aspects worthy to be investigated \cite{31,32,33}. The \textit{Hawking} temperature of Eq. (\ref{fr}) is given by
\begin{eqnarray}
T_H=\frac{1}{4\pi} \partial_{r} f(r)\big|_{r=r_h}=\frac{-Q-(\bar{l}-1) r_h^2}{4(\bar{l}-1)^2 \pi  r_h^3}\label{TH},
\end{eqnarray}
where $r_h$ is referred to horizon radius and is calculated from $f(r)\big|_{r=r_h}=0$. In addition, the the respective entropy reads
\begin{eqnarray}
S=\pi r_h^2,
\end{eqnarray}
so that the heat capacity as
\begin{eqnarray}
C_v = T_H \frac{\partial S}{\partial T_H}=T_H\frac{\partial S / \partial r_h}{\partial T_H/ \partial r_h}.
\end{eqnarray}
In order to provide a better interpretation of our results, the illustration of $T_H$ versus entropy is shown in Fig \ref{fig:TS}. 
Here, we also make a comparison with the case of $Q=0.0$. The heat capacity curve due changing horizon radius is shown in Fig. \ref{fig:Cv}.
Notice that, by increasing parameter $\bar{l}$, the phase transition occurs in small values of horizon radii. 

%%%%%%%%%%%%%%%%%%%%%%%%%%%%%%%%%%%%%%%%%%%%%%%%%%%%%%%%%%%%%%%%%%%%%%%%%%%%%%%%%%%%%%%%%%%%%%%%%%%%%%%%%%%%%%%%%%%%%%%%%%%%%%%%%%%%%%%%%%%%%%%%%%%%%%%%%%%%%%%%%%%%%%%%%%%%%%%%%%%%%%%%%%%%%%%%%%%%%%%%%%%%%%%%%%%%%%%%%%%%%%%%%%%%%%%%%%%%%%%%%%%%%%%%%%%%%%%%%%%%%%%%%%%%%%%%%%%%%%%%%%%%%%%%%%%%%%%%%%%%%%%%%%%%%%%%%%%%%%%%%%%%%%%%%%%%%%%%%%%%%%%%%%%%%%%%%%%%%%%%%%%%%%%%%%%%%%%%%%%%%%%%%%%%%%%%%%%%%%%%%%%%%%%%%%%%%%%%%%%%%%%%%%%%%%%%%%%%%%%%%%%%%%%%%%%%%%%%%%%%%%%%%%%%%%%%%%%%%%%%%%%%%%%%%%%%%%%%%%%%%%%%%%%%%%%%%%%%%%%%

\section{\textit{Greybody} Bounds, Emission Power, and Partial Absorption Cross-Section}\label{Sec4}

\begin{figure}[ht!]
\centering
	\begin{subfigure}[b]{0.5\textwidth}
	\centering
	\includegraphics[scale = 0.48]{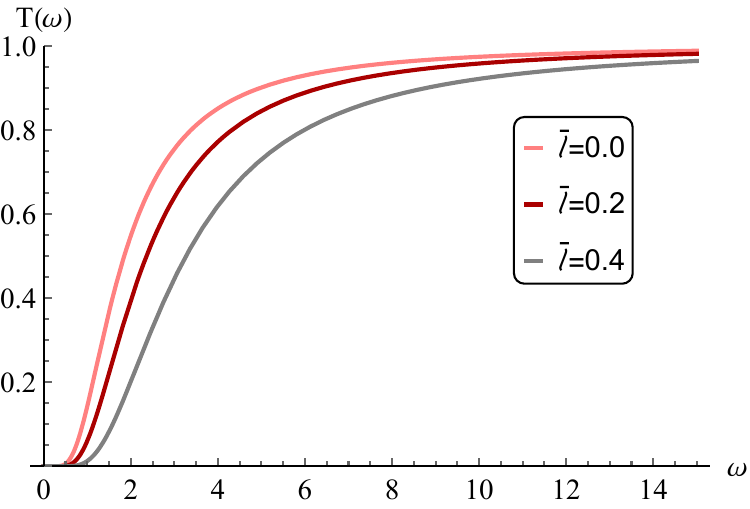} \hspace{-0.2cm}
%	\caption{}
    \end{subfigure}%
\hfill
\caption{\textit{Greybody} bounds for different $\bar{l}$s. $l=2$ and other parameters are set as $M=1,\,Q=0.01,\,s=0$.}
\label{fig:gbb}
\end{figure}

The \textit{greybody} bounds show the radiation of the particles emitted from the BH. The maximum value of \textit{greybody} bound belongs to the blackbody and is equal to one. For other issues, its upper limit is calculated from from \cite{16,20,21,22}
\begin{eqnarray}
T_l(\omega)\geq \text{sech}^2 \left(\frac{1}{2\omega}\int_{r_h}^{\infty}V(r) \frac{\mathrm{d}r}{f(r)}\right)\label{gbb},
\end{eqnarray}
that $\omega$ demonstrates the frequency and $V(r)$ referred to the effective potential and is given by
\begin{eqnarray}
V(r)=f(r) \left(\frac{(1-s^2)\partial_r f(r)}{r}+\frac{(l+1) l}{r^2}\right)\label{potential},
\end{eqnarray}
where $l$ represents the angular momentum and $s$ shows the type of perturbation, e.g., $s=0$, $s=1$ and $s=2$ are referred to the scalar perturbation, electromagnetic and tensorial perturbations, respectively. Fig. \ref{fig:gbb} illustrates the \textit{greybody} curve due changing the frequency.
It is evident that the probability of transmission for a particular $Q$, decreases by increasing parameter $\bar{l}$. For a BH that is in thermal equilibrium with its surroundings, the temperature remains constant for large values of $\omega$. The $l$th mode emitted power of such system is given by \cite{16,20}
\begin{eqnarray}
P_{l}(\omega)=\frac{A}{8\pi^2}T_l(\omega)\frac{\omega^3}{\exp(\omega / T_H)-1},
\end{eqnarray}
where $A$ represents the surface area and $T_H$ shows \textit{Hawking} temperature.
\begin{figure}[ht!]
\centering
	\begin{subfigure}[b]{0.5\textwidth}
	\centering
	\includegraphics[scale = 0.48]{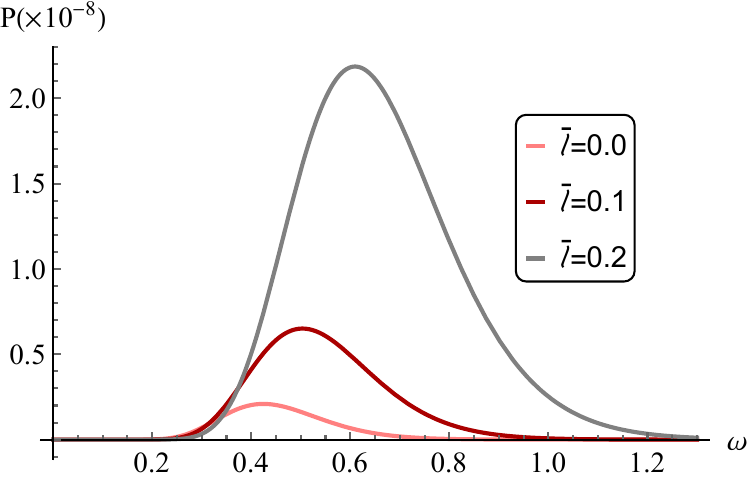} \hspace{-0.2cm}\\
%	\caption{}
    \end{subfigure}%
\hfill
\caption{The emitted power for three $\bar{l}$s. $l=2$ and other parameters are set as $M=1,\,Q=0.01,\,s=0$.}
\label{fig:gbbpower}
\end{figure}
As shown in Fig. \ref{fig:gbbpower}, by varying parameter $\bar{l}$, the maximum value of the emitted power and also the position of the maximum frequency are shifted. The absorption cross section by using the transmission probability which is given in Eq. (\ref{gbb}) is calculated from \cite{27,28}
\begin{eqnarray}
\sigma_{abs}^l =\frac{\pi (2l+1)}{\omega^2} |T_l(\omega)|^2.
\end{eqnarray}
\begin{figure}[H]
\centering
	\begin{subfigure}[b]{0.5\textwidth}
	\centering
	\includegraphics[scale = 0.48]{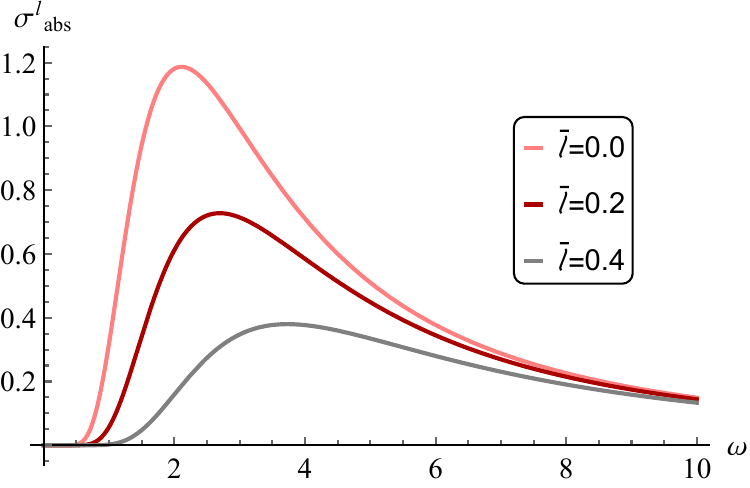} \hspace{-0.2cm}
%	\caption{}
    \end{subfigure}%
\hfill
\caption{The absorption cross--section curve for different $\bar{l}$'s. $l=2$ and other parameters are set as $M=1,\,Q=0.01,\,s=0$.}
\label{fig:gbb_sigma}
\end{figure}
Fig. \ref{fig:gbb_sigma} indicates that by decreasing parameter $\bar{l}$, the maximum value of absorption and its relevant frequency is shifted.

%%%%%%%%%%%%%%%%%%%%%%%%%%%%%%%%%%%%%%%%%%%%%%%%%%%%%%%%%%%%%%%%%%%%%%%%%%%%%%%%%%%%%%%%%%%%%%%%%%%%%%%%%%%%%%%%%%%%%%%%%%%%%%%%%%%%%%%%%%%%%%%%%%%%%%%%%%%%%%%%%%%%%%%%%%%%%%%%%%%%%%%%%%%%%%%%%%%%%%%%%%%%%%%%%%%%%%%%%%%%%%%%%%%%%%%%%%%%%%%%%%%%%%%%%%%%%%%%%%%%%%%%%%%%%%%%%%%%%%%%%%%%%%%%%%%%%%%%%%%%%%%%%%%%%%%%%%%%%%%%%%%%%%%%%%%%%%%%%%%%%%%%%%%%%%%%%%%%%%%%%%%%%%%%%%%%%%%%%%%%%%%%%%%%%%%%%%%%%%%%%%%%%%%%%%%%%%%%%%%%%%%%%%%%%%%%%%%%%%%%%%%%%%%%%%%%%%%%%%%%%%%%%%%%%%%%%%%%%%%%%%%%%%%%%%%%%%%%%%%%%%%%%%%%%%%%%%%%%%%%

\section{Topological Charge}\label{Sec5}
%Considering $h(r)$ in \ref{ds2} as
%\begin{eqnarray}
%h(r)=r^2,
%\end{eqnarray}
A potential function could be defined as \cite{23,24}
\begin{eqnarray}
H(r,\theta)=\frac{1}{\sin\theta}\sqrt{\frac{f(r)}{h(r)}},
\end{eqnarray}
which is used to investigate the topology of the BH. In order to find the topological phase transition of light rings, a vector field is used as \cite{23}
\begin{eqnarray}
\phi_{r}=\sqrt{f(r)} \partial_r H(r,\theta),\;
\phi_{\theta}=\frac{1}{\sqrt{h(r)}}\partial_{\theta} H(r,\theta),
\end{eqnarray}
which can be normalized as
\begin{eqnarray}
n_r=\frac{\phi_{r}}{||\phi||},\;n_{\theta}=\frac{\phi_{\theta}}{||\phi||}\label{unitv}.
\end{eqnarray}
The vector space of form $(n_r,n_\theta)$ is employed to show topological phase transition. The direction of the vectors and their change in direction is used to investigate the phase transition in the system. It is worh commenting that the topology of nonlinearly charged black hole chemistry via massive gravity was recently addressed in the literature \cite{zhang2023topology}. In next section, the topological charge of the phase transitions will be determined.

%%%%%%%%%%%%%%%%%%%%%%%%%%%%%%%%%%%%%%%%%%%%%%%%%%%%%%%%%%%%%%%%%%%%%%%%%%%%%%%%%%%%%%%%%%%%%%%%%%%%%%%%%%%%%%%%%%%%%%%%%%%%%%%%%%%%%%%%%%%%%%%%%%%%%%%%%%%%%%%%%%%%%%%%%%%%%%%%%%%%%%%%%%%%%%%%%%%%%%%%%%%%%%%%%%%%%%%%%%%%%%%%%%%%%%%%%%%%%%%%%%%%%%%%%%%%%%%%%%%%%%%%%%%%%%%%%%%%%%%%%%%%%%%%%%%%%%%%%%%%%%%%%%%%%%%%%%%%%%%%%%%%%%%%%%%%%%%%%%%%%%%%%%%%%%%%%%%%%%%%%%%%%%%%%%%%%%%%%%%%%%%%%%%%%%%%%%%%%%%%%%%%%%%%%%%%%%%%%%%%%%%%%%%%%%%%%%%%%%%%%%%%%%%%%%%%%%%%%%%%%%%%%%%%%%%%%%%%%%%%%%%%%%%%%%%%%%%%%%%%%%%%%%%%%%%%%%%%%%%%

\section{Deflection Angle of Topological Charge}\label{Sec6}

\begin{figure}[ht!]
\centering
	\begin{subfigure}[b]{0.5\textwidth}
	\centering
	\includegraphics[scale = 0.48]{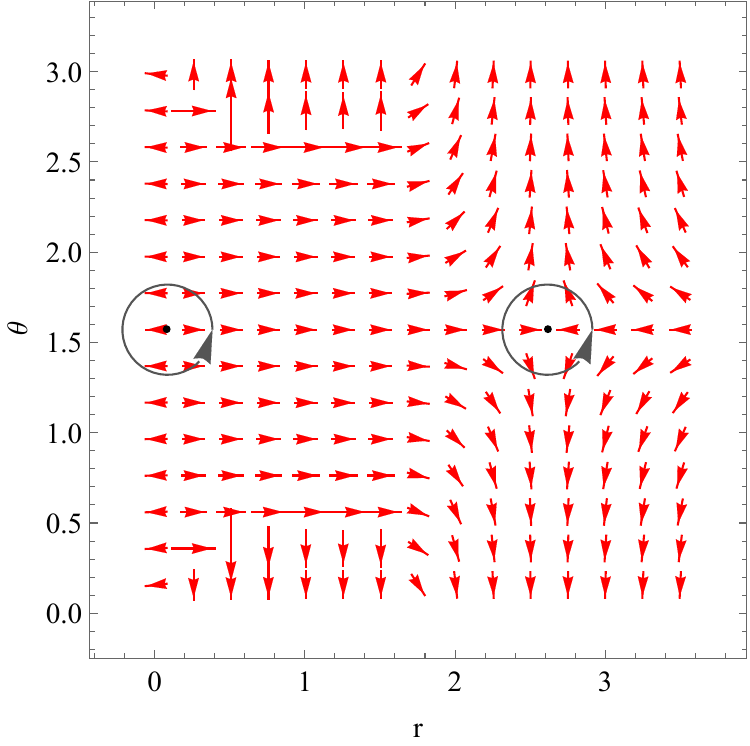} \hspace{-0.2cm}\\
	\caption{Parameters are set as $Q=0.1,\,\bar{l}=0.1$. At $r = 2.61502,\, w=-1$ and at $r = 0.08498,\, w=1$.}
	\hspace{-0.2cm}\\
    \end{subfigure}%
    \hfill
	\begin{subfigure}[b]{0.5\textwidth}
	\centering
	\includegraphics[scale = 0.48]{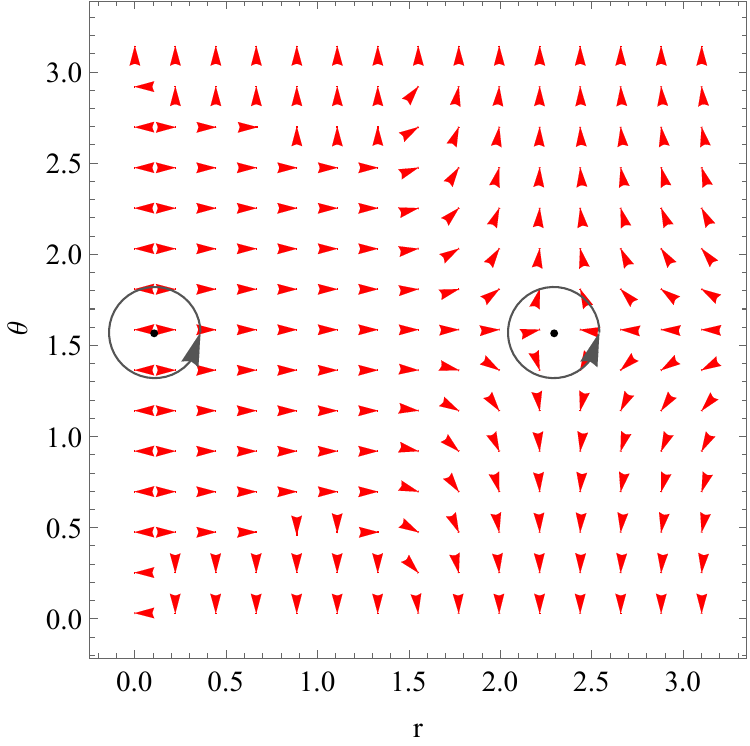} \hspace{-0.2cm}\\
	\caption{Parameters are set as $Q=0.1,\,\Bar{l}=0.2$. At $r = 2.29087,\, w=-1$ and at $r = 0.10913,\, w=1$}
	\hspace{-0.2cm}\\
    \end{subfigure}%
    \hfill
\caption{The vector space is shown. The whole charge for such system is $\mathfrak{Q}=0$.}
\label{fig:phasetr}
\end{figure}

According to Refs. \cite{24,25}, assuming the deflected angle of topological charge as
\begin{eqnarray}
\Omega=\arctan \frac{\phi_\theta}{\phi_r}=\arctan\frac{n_\theta}{n_r},
\end{eqnarray}
we have
\begin{eqnarray}
d \Omega=\frac{n_r \partial n_{\theta}-n_{\theta} \partial n_r}{n_{r}^2+n_{\theta}^{2}},
\end{eqnarray}
Due to the metric form, $d \Omega$ has one or more poles at $\theta = \pi/2$ and photon radii. For the sake of avoiding singularities, we can use the following parameters
\begin{align}
\nonumber
r &= a \cos\vartheta +r_{p},\\
\theta &= b \sin\vartheta+\frac{\pi}{2},
\end{align}
and the topological charge can be calculated from
\begin{eqnarray}
\mathfrak{Q}=\frac{\Delta\Omega}{2\pi}=\frac{1}{2\pi}\oint_C n_r \partial_\vartheta n_\theta - n_\theta \partial_\vartheta n_r,
\end{eqnarray}
where the poles of $\Delta\Omega$ occur at $(r_{ph},\pi/2)$ and $\mathfrak{Q}$ takes $0,\,\pm 1$. The illustration of vector space $(n_r,\,n_{\theta})$ is shown in Fig. \ref{fig:phasetr}.
The phase transition occurs in the solution of the Eq. (\ref{rph}) and at the point $(r_{ph},\pi/2)$. As Eq. (\ref{rph}) have two solutions, two phase transitions occur and the entire topological charge becomes zero.

%%%%%%%%%%%%%%%%%%%%%%%%%%%%%%%%%%%%%%%%%%%%%%%%%%%%%%%%%%%%%%%%%%%%%%%%%%%%%%%%%%%%%%%%%%%%%%%%%%%%%%%%%%%%%%%%%%%%%%%%%%%%%%%%%%%%%%%%%%%%%%%%%%%%%%%%%%%%%%%%%%%%%%%%%%%%%%%%%%%%%%%%%%%%%%%%%%%%%%%%%%%%%%%%%%%%%%%%%%%%%%%%%%%%%%%%%%%%%%%%%%%%%%%%%%%%%%%%%%%%%%%%%%%%%%%%%%%%%%%%%%%%%%%%%%%%%%%%%%%%%%%%%%%%%%%%%%%%%%%%%%%%%%%%%%%%%%%%%%%%%%%%%%%%%%%%%%%%%%%%%%%%%%%%%%%%%%%%%%%%%%%%%%%%%%%%%%%%%%%%%%%%%%%%%%%%%%%%%%%%%%%%%%%%%%%%%%%%%%%%%%%%%%%%%%%%%%%%%%%%%%%%%%%%%%%%%%%%%%%%%%%%%%%%%%%%%%%%%%%%%%%%%%%%%%%%%%%%%%%%

\section{Phase Transition in Temperature and Free Energy}\label{Sec7}

\begin{figure}[ht!]
\centering
	\begin{subfigure}[b]{0.5\textwidth}
	\centering
	\includegraphics[scale = 0.48]{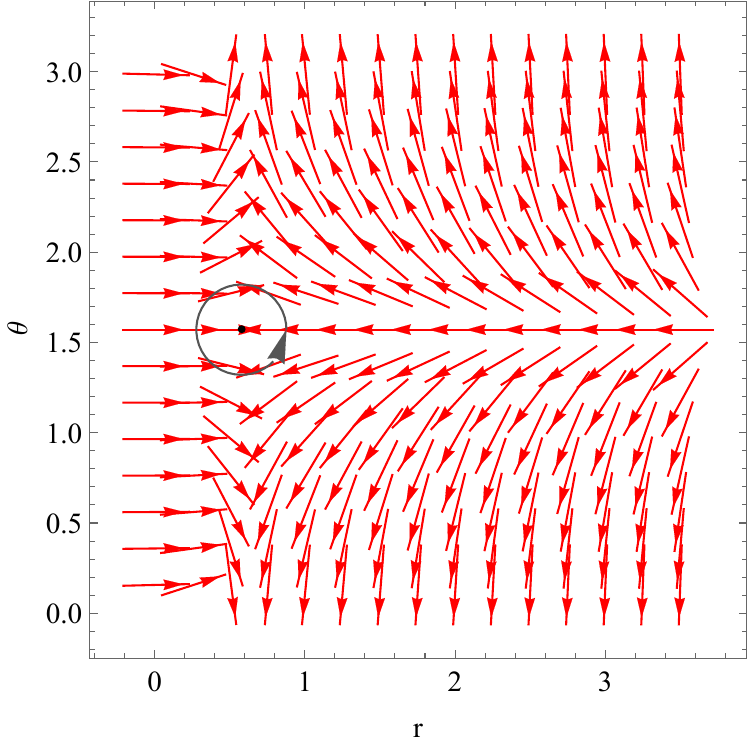} \hspace{-0.2cm}\\
%	\caption{}
    \end{subfigure}%
\hfill
\caption{Temperature topological charge for $Q=0.1$ and $\bar{l}=0.1$ at $r=0.57735$. In this case $\mathfrak{Q}=-1$.}
\label{fig:PhaseTemp}
\end{figure}

Using \textit{Hawking} temperature, which is calculated in Eq. (\ref{TH}), a field can be defined as
\begin{eqnarray}\nonumber
\Phi =\frac{1}{\sin (\theta)} T_H=-\frac{\csc (\theta ) \left((\bar{l}-1) r_h^2+Q\right)}{4 \pi  (\bar{l}-1)^2 r_h^3}.
\end{eqnarray}
The associated vectors are shown below
\begin{eqnarray}
\varphi_r=\partial_{r_h}\Phi,\;\varphi_{\theta}=\partial_{\theta}\Phi,
\end{eqnarray}
and unit vectors could be calculated by using Eq. (\ref{unitv}). Fig. \ref{fig:PhaseTemp} shows the topological charge of the system which is caused by temperature. In this case, the whole topological charge of the system is $\mathfrak{Q}=-1$.

\begin{figure}[ht!]
\centering
	\begin{subfigure}[b]{0.5\textwidth}
	\centering
	\includegraphics[scale = 0.5]{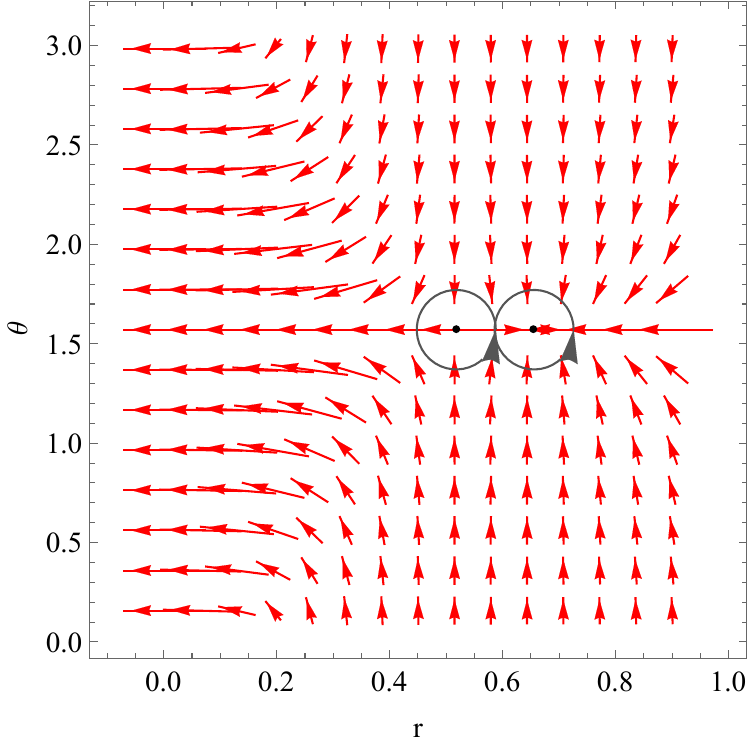} \hspace{-0.2cm}
    \end{subfigure}%
\hfill
\caption{Free energy topological charge for $Q=0.1,$, $\bar{l}=0.1$ and $\tau=10$ at $r=0.51785, w=1$, and $r=0.65566, w=-1$. In this case the topological charge is $\mathfrak{Q}=0$.}
\label{fig:PhaseHelm}
\end{figure}

In addition, the Helmholtz free energy reads
\begin{align}
\nonumber
\mathcal{F}=M(r_h)- \frac{S}{\tau} =\frac{Q}{2 (\bar{l}-1)^2 r_h}+ \frac{r_h}{(1- \bar{l})}-\frac{\pi r_h^2}{\tau},
\end{align}
where $\tau =1/T_H$. In this case, a new field and its related vectors are derived as follows
\begin{eqnarray}
\phi_r=\partial_{r_h}\mathcal{F},\;\phi_{\theta}=-\cot\theta\csc\theta.
\end{eqnarray}
Fig. \ref{fig:PhaseHelm} represents the topological charged that is derived from free energy. Naturally, the total charge of such system equals $\mathfrak{Q}=0$.

%%%%%%%%%%%%%%%%%%%%%%%%%%%%%%%%%%%%%%%%%%%%%%%%%%%%%%%%%%%%%%%%%%%%%%%%%%%%%%%%%%%%%%%%%%%%%%%%%%%%%%%%%%%%%%%%%%%%%%%%%%%%%%%%%%%%%%%%%%%%%%%%%%%%%%%%%%%%%%%%%%%%%%%%%%%%%%%%%%%%%%%%%%%%%%%%%%%%%%%%%%%%%%%%%%%%%%%%%%%%%%%%%%%%%%%%%%%%%%%%%%%%%%%%%%%%%%%%%%%%%%%%%%%%%%%%%%%%%%%%%%%%%%%%%%%%%%%%%%%%%%%%%%%%%%%%%%%%%%%%%%%%%%%%%%%%%%%%%%%%%%%%%%%%%%%%%%%%%%%%%%%%%%%%%%%%%%%%%%%%%%%%%%%%%%%%%%%%%%%%%%%%%%%%%%%%%%%%%%%%%%%%%%%%%%%%%%%%%%%%%%%%%%%%%%%%%%%%%%%%%%%%%%%%%%%%%%%%%%%%%%%%%%%%%%%%%%%%%%%%%%%%%%%%%%%%%%%%%%%%

\section{Conclusion}\label{Sec9}

In this work, we investigated the impact of an antisymmetric Kalb--Ramond tensor field, which leads to the spontaneous breaking of Lorentz symmetry, on the properties of a charged black hole. Our study focused on examining the shadow radius, \textit{greybody} bound, relevant absorption and emission power, heat capacity, topological charge, and optical appearance. This research aims to fill a gap in the literature and provide more information about the implications of this Lorentz symmetry-breaking scenario.

Initially, we calculated the shadow radius. For the special case \( M = 1 \), we determined the lower and upper bounds of the parameter \( \bar{l} \) for three different values of the parameter \( Q \). We also investigated features that deviate from the standard charged BH solutions, such as the \textit{greybody} bound and its relevant absorption cross-section. For \( Q = 0.01 \), an increase in the parameter \( \bar{l} \) lowered the curves, while the emission power curve moved higher with the increment of \( \bar{l} \).

Notably, we explored the topological charge and the associated topological phase transitions that occur in this scenario. By analyzing the metric, temperature, and free energy, we identified the topological charge of the system and the corresponding phase transitions.

As a further perspective, we can apply similar calculations to other configurations of black hole solutions involving the Kalb-Ramond field, as presented in Ref. \cite{19}. These and other ideas are currently under development.

%%%%%%%%%%%%%%%%%%%%%%%%%%%%%%%%%%%%%%%%%%%%%%%%%%%%%%%%%%%%%%%%%%%%%%%%%%%%%%%%%%%%%%%%%%%%%%%%%%%%%%%%%%%%%%%%%%%%%%%%%%%%%%%%%%%%%%%%%%%%%%%%%%%%%%%%%%%%%%%%%%%%%%%%%%%%%%%%%%%%%%%%%%%%%%%%%%%%%%%%%%%%%%%%%%%%%%%%%%%%%%%%%%%%%%%%%%%%%%%%%%%%%%%%%%%%%%%%%%%%%%%%%%%%%%%%%%%%%%%%%%%%%%%%%%%%%%%%%%%%%%%%%%%%%%%%%%%%%%%%%%%%%%%%%%%%%%%%%%%%%%%%%%%%%%%%%%%%%%%%%%%%%%%%%%%%%%%%%%%%%%%%%%%%%%%%%%%%%%%%%%%%%%%%%%%%%%%%%%%%%%%%%%%%%%%%%%%%%%%%%%%%%%%%%%%%%%%%%%%%%%%%%%%%%%%%%%%%%%%%%%%%%%%%%%%%%%%%%%%%%%%%%%%%%%%%%%%%%%%%

\section*{Acknowledgments}
\hspace{0.5cm}

A. A. Araújo Filho would like to thank Fundação de Apoio à Pesquisa do Estado da Paraíba (FAPESQ) and Conselho Nacional de Desenvolvimento Cientíıfico e Tecnológico (CNPq)  -- [150891/2023-7] for the financial support.

\bibliographystyle{ieeetr}
\bibliography{main}

\end{document}